\documentclass[prd,twocolumn,tightenlines,superscriptaddress,nofootinbib,
showpacs]{revtex4}
\usepackage{amssymb,latexsym}
\usepackage{amsmath,amsbsy,bbm}
\usepackage{epsfig,bm}
\usepackage{graphicx,comment}
\usepackage{color}
\newcommand{\p}{\partial}

\begin{document} 

\title{Monte Carlo determination of the low-energy constants for 
a two-dimensional spin-1 Heisenberg model with spatial anisotropy}

\author{F.-J. Jiang}
\email[]{fjjiang@ntnu.edu.tw}
\affiliation{Department of Physics, National Taiwan Normal University, 
88, Sec.4, Ting-Chou Rd., Taipei 116, Taiwan}

\vspace{-2cm}
  
\begin{abstract}

The low-energy constants, namely the spin stiffness $\rho_s$, the staggered
magnetization density ${\cal M}_s$ per area, and the spinwave velocity 
$c$ of the two-dimensional (2D) spin-1 Heisenberg model on the square and
rectangular lattices are determined using the first principles Monte Carlo 
method. In particular, the studied models have antiferromagnetic couplings
$J_1$ and $J_2$ in the spatial 1- and 2-directions, respectively. 
For each considered $J_2/J_1$, the aspect ratio of the corresponding 
linear box sizes $L_2/L_1$ used in the simulations
is adjusted so that the squares of the two spatial winding numbers take
the same values. In addition,
the relevant finite-volume 
and -temperature predictions from magnon chiral perturbation theory are employed 
in extracting the numerical values of these low-energy constants.
Our results of $\rho_{s1}$ are in quantitative agreement 
with those obtained by the series expansion method over a broad range of $J_2/J_1$. 
This in turn provides  
convincing numerical evidence for the quantitative correctness of our 
approach. The ${\cal M}_s$ and $c$ presented here for 
the spatially anisotropic models are new and can be used as benchmarks for future related 
studies. 

\end{abstract}
%\pacs{12.39.Fe, 75.10.Jm, 75.40.Mg, 75.50.Ee}

\maketitle
\vskip-0.5cm

\section{Introduction}

During the last three decades, the two-dimensional (2D) spin-1/2 Heisenberg 
model and its generalizations have been investigated in great detail
both analytically and numerically 
\cite{Cana92,Oit941,Oit942,Wie94,San97,Bea96,Jia08,Jia11.2}.
This is because these models are regarded as the relevant models 
for the 2D quantum antiferromagnets. Furthermore,
although the phase diagram of high temperature cuprate superconductors
is not well understood, it is generally believed that these cuprate
superconductors may be obtained by doping the quantum antiferromagnetic 
insulators with charge carriers. As a result, research related to these 
models is still very active even today.

\begin{figure}
\begin{center}
\includegraphics[width=0.35\textwidth]{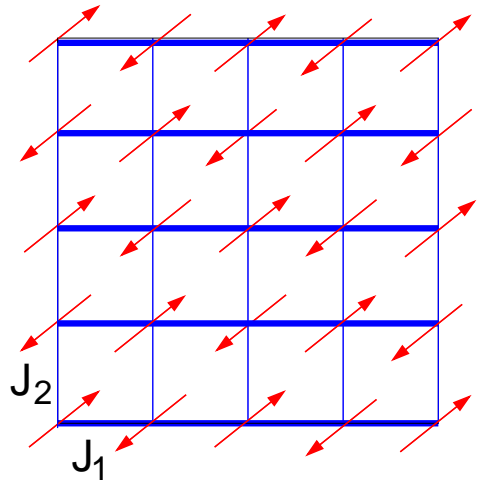}
\end{center}
\caption{The two-dimensional (2D) spatially anisotropic spin-1 Heisenberg 
model on the square and rectangular lattices investigated
in this study.}
\label{fig0}
\end{figure}

In addition to the spin-1/2 Heisenberg model, higher spin antiferromagnets,
in particular the spin-1 Heisenberg model, 
are of theoretical interest due to the fact that they are
relevant in explaining experimental results of real materials as well \cite{Ren87,Yam95,Mon96,Hon97,Hag98,Hon98,Uch99,Nar01}.
For example, NDMAZ, NENP, and PbNi$_2$V$_2$O$_8$ are found to be spin-1 
quasi-one-dimensional antiferromagnets. Besides the simplest type of these models which have
spatially isotropic couplings, the spatially anisotropic Heisenberg models
are also studied thoroughly \cite{Hal831,Hal832,Aff87,Sin88,Sak89,Egg94,Kat94,Yam951,Kit96,Koh98,Kog00,Kim00,Matsumoto02,Wan05,Wen09,Pardini08,Jia09}.  
In particular, these generalized models are frequently used as a route for studying quantum phase 
transitions. Moreover, the spatially anisotropic models
are important in understanding experimental data. Two notable 
examples are the 2D 
spin-1/2 Heisenberg model with antiferromagnetic couplings $J_1$ and $J_2$ 
\footnote{In this study a physical 
quantity with a subscript $i$ refers to its value in the spatial $i$-direction.}
on the square and rectangular lattices as depicted in fig.~\ref{fig0}, and the three-dimensional
(3D) quantum antiferromagnet with a ladder pattern of spatial anisotropy on
a cubic lattice. The former is argued to be relevant to 
the newly discovered pinning effects of 
the electronic liquid crystal in the underdoped cuprate superconductor 
YBa$_2$Cu$_3$O$_{6.45}$ \cite{Hinkov2007,Hinkov2008}, and the latter 
is considered to be the right model for explaining the phase diagram
of TlCuCl$_3$ under pressure \cite{Rue03,Rue08,Jin12,Oit12,Kao13,Mer14}. To conclude, despite their simplicity, 
the spatially anisotropic Heisenberg models are among the most
important and frequently studied systems in condensed matter physics.

Among the spatially anisotropic Heisenberg models with quantum spin,
the one shown in fig. 1 is particularly special.
For this model one sees clearly that as the magnitude of the
ratio of couplings $J_2/J_1$ decreases, the system will eventually become
decoupled one-dimensional (1D) antiferromagnetic chains (This
takes place when $J_2/J_1 = 0$).
One intriguing physics to explore for this spin-1/2 model is to examine whether a phase
transition, between the antiferromagnetic and dimerized
phases, occurs before one reaches the extreme case $J_2/J_1 = 0$. 
Analytic (and some numerical) 
evidence indicates that for the model of fig.~1 with quantum spin, 
the long-range antiferromagnetic order is destroyed
only for infinitesimal $J_2/J_1$ \cite{Aff94,Aff96}. As a result, 
to study whether a phase transition
appears at a particular value of $J_2/J_1 > 0$ using unbiased quantum Monte Carlo simulations is subtle. 
Indeed, as suggested in Refs.~\cite{Matsumoto02,San99}, square 
lattice is not the appropriate lattice geometry for studying this model 
and rectangular lattice should be used instead. Furthermore, to capture the
2D characters of the model, the ratio of linear lattice sizes $L_2/L_1$ needs 
to be adjusted individually for each considered $J_2/J_1$. Even so, one must 
carry out careful investigation for the relevant observables so that the 
correct physics is obtained.
For example, to make sure that the extrapolating results are reliable, 
for every $J_2/J_1$ several ratios of $L_2/L_1$ may be needed in the 
calculations.

For the spin-1/2 model depicted in fig.~1, 
one proposed quantitative approach to study the 2D ground state properties
for every considered $J_2/J_1$   
is to adjust the corresponding ratio $L_2/L_1$ so that the spatial winding 
number squared in the 1- and 2-directions, namely $\langle W_1^2\rangle$ and 
$\langle W_2^2\rangle$ take the same values. With such a method
the 2D characters of the model will not be lost and can be
obtained unbiasedly.  
This idea and similar ones have been used to study the spin-1/2 model 
(of fig.~1) \cite{Jia09} 
as well as quantum phase transitions of 2D dimerized quantum spin models
\cite{Jia12,Shi13}.  

The motivation of our study presented here is to provide
a more convincing numerical result to support the quantitative correctness of
the method of requiring $\langle W_1^2\rangle = \langle W_2^2\rangle$
in the simulations. Indeed, for the model in fig.~1 with quantum spin, 
Monte Carlo and series expansion results
of the observable spin stiffness in the 1-direction agree with each other only for $J_2/J_1 > 0.4$ \cite{Jia09}.
Therefore it is desirable to carry out a more detailed investigation so that 
the validity
of this method can be justified.
Notice it is known that the quantum phase transition associated with 
dimerization for the spin-1 model with the same spatial anisotropy takes place at 
a finite value of $J_2/J_1$ \cite{Matsumoto02}. Hence 
simulating the spin-1 model with 
the same spatial anisotropy shown in fig.~1 provides a great opportunity
to further examine the validity of this method.     
Indeed as we will demonstrate later, our Monte Carlo results of $\rho_{s1}$ 
for the spin-1 model agree quantitatively with those determined 
by the series expansion method from $J_2/J_1 = 1$ down to $J_2/J_1 \sim 0.08$ \cite{Pardini08}. 
Consequently our investigation
gives convincing evidence for the correctness of this method. 

Notice the spatially anisotropic quantum Heisenberg model of fig.~\ref{fig0}
and its 1D limit are two completely different systems \cite{Cha89,Neu89,Has90,Has91,Egg94,Has93}.  
Therefore an unconventional behavior is likely to appear as one approaches the 1D limit from the 2D 
model. Our study paves a way to investigate the novel
phenomena of dimension crossover from a 2D system to its 1D limit.
We would like to point out that even with the method of adjusting the aspect ratio
$L_2/L_1$ in the calculations so that the two spatial winding numbers squared reach the same numbers,
the determination of ground state bulk properties requires the simulations being conducted at very low temperatures.
Such zero temperature calculations are computationally demanding.
Therefore instead of performing the simulations at very low temperatures and using the conventional approach of 
fitting the data with polynomials of the relevant parameters, 
we carry out the calculations at finite temperatures and employ the relevant predictions from magnon chiral 
perturbation theory (m$\chi$PT) to extract the bulk properties of the 
considered model.  
Later we will briefly argue that under certain circumstances, this approach seems to be more efficient than the conventional one. Indeed, 
for each considered $J_2/J_1$, a few tens of the corresponding data points can be described quantitatively by two 
equations with only three unknown parameters, and particularly,
we are able to arrive at high precision results with moderate computing resources. 
Finally, another remarkable finding in our investigation is that,
the m$\chi$PT can also be used to examine unambiguously the presence of
the long-range antiferromagnetic order. This will be explained in more detail 
in the relevant section.

This paper is organized as follows. First, after the introduction,
the spatially anisotropic spin-1 Heisenberg model and the 
observables considered in this work are detailed. Furthermore,
the m$\chi$PT is briefly introduced 
and some of its predictions relevant to our study are also listed. 
We then present the data as well as the resulting
numerical results based on these data. Finally a section is devoted to conclude our 
investigation.

\section{Microscopic Models and Corresponding Observables}
%\label{micro}
The 2D spin-1 Heisenberg model we consider in this study is defined by the 
Hamilton operator
\begin{eqnarray}
\label{hamilton}
H = \sum_{x} \Big[\,J_{1}\vec S_x \cdot \vec S_{x+\hat{1}a}+J_{2} 
\vec S_x \cdot \vec S_{x+\hat{2}a}\,\Big],
\end{eqnarray}
and is depicted by fig.~1. In Eq.~(1), $\hat{1}$ and $\hat{2}$ refer to the two spatial unit-vectors and
$a$ is the lattice spacing. In addition, $J_{1}$ and $J_{2}$ are the antiferromagnetic couplings in the $1$- and $2$-direction, respectively.
Finally the $\vec{S_i}$ shown above is a spin-1 operator at site $i$. 

A physical quantity measured in our simulations 
is the staggered susceptibility $\chi_s$, which is given by
\begin{eqnarray}
\label{defstagg}
\chi_s 
&=& \frac{1}{L_{1} L_{2}} \int_0^\beta dt \ \frac{1}{Z} 
\mbox{Tr}[M^3_s(0) M^3_s(t) \exp(- \beta H)].
\end{eqnarray}
Here $\beta$ is the inverse temperature, $L_{1}$ and $L_{2}$ are the spatial box
sizes in the $1$- and $2$-direction, respectively, and 
$Z = \mbox{Tr}\exp(- \beta H)$
is the partition function. Furthermore, $M^3_s$ appearing in Eq.~(2) is 
the third component of the staggered magnetization 
$\vec M_s = \sum_x (-1)^{x_1+x_2} \vec S_x$.
Another observables considered here is the uniform susceptibility $\chi_u$,
which takes the form
\begin{eqnarray}
\label{defuniform}
\chi_u 
&=& \frac{1}{L_1 L_2} \int_0^\beta dt \ \frac{1}{Z} \mbox{Tr}[M^3(0) M^3(t)
\exp(- \beta H)].
\end{eqnarray}
Here $\vec{M} = \sum_x \vec S_x$ 
is the uniform magnetization. Both $\chi_s$ and $\chi_u$ can be measured very 
efficiently with the loop-cluster algorithm using improved estimators 
\cite{Wie94}. 

Notice both of these two observables can be expressed in quantities 
related to the clusters. In particular, $\chi_u$ is associated
with the temporal winding number $W_t = \sum_{\cal C} W_t({\cal C})$ which is the sum of winding numbers
$W_t({\cal C})$ of the loop-clusters ${\cal C}$ around the Euclidean time 
direction. Similarly, the spatial winding numbers $W_i$ for $i\in \{1,2\}$ are defined in the same 
manner. With the convention employed here,
the spin stiffness $\rho_s$ (for $J_2/J_1 = 1$)
can be obtained directly from the standard relation 
$\rho_s = \frac{3}{4\beta}\left( \langle W_1^2 \rangle + \langle W_2^2 \rangle\right)$
in the zero-temperature and infinite-volume limits. In addition,
the temporal winding number squared $\langle W_t^2\rangle$ calculated in this study is exactly 
the susceptibility $\chi$ and is related to $\chi_u$ by $\chi_u = \frac{\beta}{L_1 L_2}\langle W_t^2\rangle$.
Finally the spinwave velocity $c$ can be estimated from 
these winding numbers squared by the method detailed in Refs.~\cite{Jia11.1,Sen15,Kao13}.

\section{Low-Energy Effective Theory for Magnons}
%\label{lowen}
In this section we summarize the relevant theoretical predictions, namely
the finite-volume and -temperature expressions of $\chi_s$ and $\chi_u$
from m$\chi$PT \cite{Cha89,Neu89,Has90,Has91,Has93}.
These predictions are used to calculate the desired low-energy constants.
Due to the spontaneous breaking of the $SU(2)_s$ spin symmetry down to its 
$U(1)_s$ subgroup, the low-energy physics of antiferromagnets is governed by
two massless Goldstone bosons, the magnons. The systematic 
low-energy effective field theory for magnons is formulated in
term of the staggered magnetization. The staggered magnetization of an 
antiferromagnet is described by a unit-vector field $\vec{e}(x)$ in the 
coset space $SU(2)_s/U(1)_s = S^2$, i.e. 
$\vec e(x) = \big(e_1(x),e_2(x),e_3(x)\big)$ with $\vec e(x)^2 = 1$.
Here $x = (x_1,x_2,t)$ denotes a point in (2+1)-dimensional space-time. To 
leading order, the Euclidean magnon low-energy effective action takes the form
\begin{eqnarray}
\label{action}
S[\vec e\,] &=& \int^{L_1}_{0} dx_1 \int^{L_2}_{0} dx_2 \int^{\beta}_{0} 
\ dt \  
\left(\frac{\rho_{s1}}{2}\p_1 \vec e \cdot \p_1 \vec e \right. \nonumber \\
&&+\left. \frac{\rho_{s2}}{2}\p_2 \vec e \cdot \p_2 \vec e + 
\frac{\rho_s}{2c^2} \p_t \vec e \cdot \p_t \vec e\right),
\end{eqnarray}
where the index $i \in \{1,2\}$ labels the two spatial directions and $t$ 
refers to the Euclidean time-direction. The temporal spin stiffness $\rho_s$
is given by $\rho_s=\sqrt{\rho_{s1}\rho_{s2}}$, where $\rho_{s1}$ 
and $\rho_{s2}$ are the spin stiffness in the 
spatial directions. Finally, the parameter $c$ in Eq.~(\ref{action}) 
is the spinwave velocity. Notice the physical quantities, namely 
$\rho_s$, $\rho_{s1}$, $\rho_{s2}$, and c appearing inside
Eq.~(\ref{action}) are the bulk ones \cite{Has93}. By introducing $x'_{1} = (\rho_{s2}/\rho_{s1})^{1/4} x_{1}$ 
and $x'_{2} = (\rho_{s1}/\rho_{s2})^{1/4} x_{2}$, Eq.~(\ref{action}) can be 
rewritten as
\begin{eqnarray}
\label{action1}
S[\vec e\,] &=& \int^{L'_{1}}_{0} dx'_1 \int^{L'_{2}}_{0} dx'_2 
\int^{\beta}_{0} dt \  
\frac{\rho_s}{2}\Big(\p'_i \vec{e} \cdot \p'_i \vec{e} \nonumber \\ 
&& + \frac{1}{c^2} \p_t \vec{e} \cdot \p_t \vec{e}\Big).
\end{eqnarray}
If one additionally requires
$L'_{1} = L'_{2} = L$, then the condition of square area is obtained.

With the Euclidean action Eq.~(\ref{action1}), 
the finite-volume and -temperature expressions 
of $\chi_s$ and $\chi_u$ in the cubical regime, where the condition
$\beta c \approx L$ is met, are calculated in Ref.~\cite{Has93}
and take the following forms
\begin{eqnarray}
\label{chiscube}
\chi_s &=& \frac{{\cal M}_s^2 L^2 \beta}{3} 
\left\{1 + 2 \frac{c}{\rho_s L l} \beta_1(l) \right. \nonumber \\
&&+\left.\left(\frac{c}{\rho_s L l}\right)^2 \left[\beta_1(l)^2 + 
3 \beta_2(l)\right] + O\left(\frac{1}{L^3}\right) \right\}
\end{eqnarray}
and 
\begin{eqnarray}
\label{chiucube}
\chi_u &=& \frac{2 \rho_s }{3 c^2} 
\left\{1 + \frac{1}{3} \frac{c}{\rho_s L l} \widetilde\beta_1(l) +
\frac{1}{3} \left(\frac{c}{\rho_s L l}\right)^2 \times \right. \nonumber \\
&&\left.
\left[\widetilde\beta_2(l) - \frac{1}{3} \widetilde\beta_1(l)^2 - 6 \psi(l)
\right]
+ O\left(\frac{1}{L^3}\right) \right\},
\end{eqnarray}
respectively. In Eqs.~(\ref{chiscube}) and (\ref{chiucube}), the 
functions $\beta_i(l)$, $\widetilde\beta_i(l)$, and $\psi(l)$, which only 
depend on $l = (\beta c/ L)^{1/3}$, are shape coefficients of the space-time box.
The explicit formulas of these shape coefficients can be found in 
Ref.~\cite{Has93}.

\section{Determination of the Low-Energy Constants}
%\label{results}
In order to determine the low-energy constants as functions of $J_2/J_1$
for the 2D spatially anisotropic spin-1 Heisenberg
model given by Eq.~(\ref{hamilton}) (and depicted in fig.~1), we have performed simulations for 
$ 0.0435 \leq J_{2}/J_{1} \leq 1.0$ with various box sizes 
using the loop algorithm \cite{Troyer08,Bau11}. 
The value of $J_2/J_1 = 0.0435$ is included
in our consideration since it is slightly below the critical point 
$(J_2/J_1)_c = 0.043648(8)$ determined in Ref.~\cite{Matsumoto02}. 
The results at $J_2/J_1 = 0.0435$ provide an opportunity to examine whether
our Monte Carlo data at $J_2/J_1 = 0.0435$ can be captured quantitatively
by the relevant predictions of m$\chi$PT.
Without loss of generality, we have set $J_1 = 1.0$ in our Monte Carlo simulations.
The cubical regime is determined by
the condition $\langle W_{1}^2 \rangle \approx \langle W_{2}^2  \rangle  \approx \langle W_{t}^2 \rangle$.
Notice Eqs.~(\ref{chiscube}) and (\ref{chiucube}) are obtained for a
(2+1)-dimensional box with equal extent in the two spatial directions.
Since $J_{2} \leq J_{1}$ in our calculations, the box sizes 
$L_1$ and $L_2$ used in the simulations must satisfy $L_2 \le L_1$ so that the 
condition $\langle W_{1}^2 \rangle \approx \langle W_{2}^2 \rangle$ can be 
fulfilled. Finally interpolation of the data points is necessary as well in order to
use Eqs.~(\ref{chiscube}) and (\ref{chiucube}).
After employing all these requirements in our calculations, 
the low-energy constants can be extracted by fitting the Monte Carlo data to
the effective field theory predictions. First of all, in the next subsection
we focus on our Monte Carlo results of the isotropic situation $J_2/J_1 =1$.

\subsection{The low-energy constants for the isotropic model}
\begin{figure}
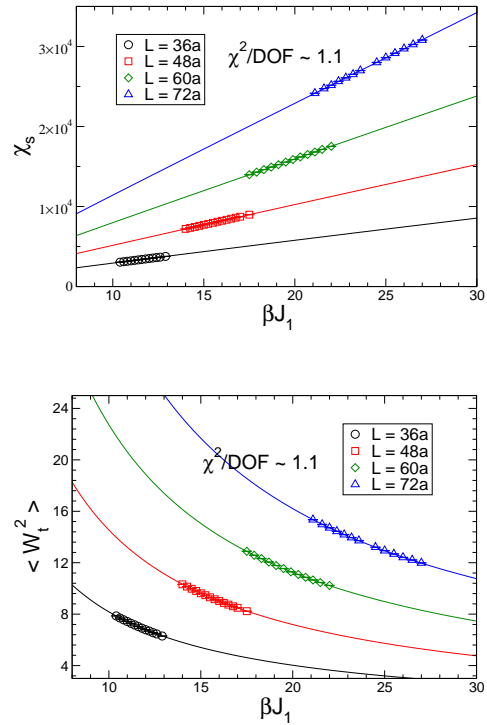

\vskip0.4cm
\begin{center}
\vbox{
\includegraphics[width=0.35\textwidth]{chi_s_J1.0.eps}\vskip0.8cm
\includegraphics[width=0.35\textwidth]{chi_u_J1.0.eps}
}
\end{center}
\caption{Results of fitting the cubical regime
data points of $\chi_s$ (top panel)
and $\langle W_t^2 \rangle$ (bottom panel) calculated at $J_2/J_1$ = $1$
to their m$\chi$PT predictions.
The solid lines are obtained using the results
from the fits.}
\label{fig1}
\end{figure}

As a first step toward a high accuracy determination of the desired low-energy 
constants, the spinwave velocity $c$ is calculated through the square of
winding numbers as suggested in Refs.~\cite{Jia11.1,Sen15}.
Specifically, for a given square lattice with linear box size $L_1$, one tunes the 
inverse temperature $\beta$ to a value $\beta^{\star}$ so that 
$\langle W_{1}^2 \rangle \approx \langle W_{2}^2  \rangle  \approx \langle W_{t}^2 \rangle$. 
Then the $c$ corresponding to this finite lattice is estimated to be $c\sim L_1/\beta^{\star}$. 
The numerical values of $c$ obtained by this method for several 
box sizes $L_1$ are given in table 1,  and a weighted average of these values of $c$ 
leads to $c = 3.0648(8)J_1a$. The result of $c$, namely $c = 3.0648(8)J_1a$ 
reached here has (much) better precision than its early known Monte Carlo results \cite{Bea99}, and is in nice agreement
with the spinwave theory estimate $c = 3.067 J_1a$ \cite{Oit942}. Furthermore, since for the isotropic case one can reach 
the condition $\langle W_{1}^2 \rangle$ = $\langle W_{2}^2\rangle$ without performing 
any interpolation, we employ Eqs.~(\ref{chiscube}) and (\ref{chiucube}) directly with a fixed 
$c = 3.0648(8)J_1a$ to simultaneously fit the uninterpolated 
data of $\chi_s$ and $\langle W_t^2\rangle $. 
The numerical values of ${\cal M}_s$ and $\rho_s$ obtained
from the fit are given by ${\cal M}_s = 0.80460(4)/a^2$ and 
$\rho_s = 0.8731(6)J_1$, respectively. In addition, the $\chi^2/{\text{DOF}}$ 
of this fit is around 1.0. 
Notice the ${\cal M}_s$ and $\rho_s$ calculated here also 
match excellently with ${\cal M}_s = 0.80426/a^2 $ and 
$2\pi \rho_s = 5.461 J_1$ determined by the spinwave theory.  
The results of the fit are shown in Fig.~\ref{fig1}.

\begin{figure}
\vskip0.4cm
\begin{center}
\includegraphics[width=0.35\textwidth]{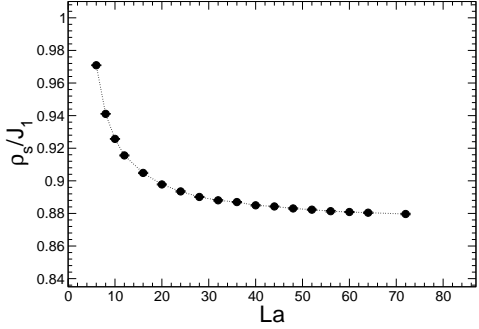}
\end{center}
\caption{The values of $\rho_s$ as a function of the box sizes $L$ for $J_2/J_1 =1.0$.
The results are obtained directly from the squares of spatial winding numbers 
and the dotted line is added to guide the eyes.}
\label{fig1.5}
\end{figure}

Besides the method of employing the predictions of m$\chi$PT, $\rho_s$ can be determined directly from the spatial 
winding numbers squared. Such a calculation of $\rho_s$ through $\langle W_1^2 \rangle$
and $\langle W_2^2 \rangle$
provides a good opportunity to verify the quantitative correctness of
m$\chi$PT. Hence we have carried out new
simulations with $J_2/J_1 = 1.0$ for several box sizes $L$ at low temperatures. 
The newly obtained $\rho_s$ data as a function of the box sizes $L$ 
is shown in Fig.~\ref{fig1.5}. Remarkably, the $\rho_s$ 
determined from the spatial winding numbers squared  
is in quantitative agreement with that calculated using 
the predictions of m$\chi$PT. For instance, by 
applying polynomials up to third (fourth) order in $1/L$ 
to the relevant data of $6 \le L \le 72$ ($6 \le L \le 72$), we 
arrive at $\rho_s = 0.8728(5)J_1$ ($\rho_s = 0.8730(10)J_1$). These values of 
$\rho_s$ match nicely with those we calculated earlier using the related 
formulas from m$\chi$PT. 
To extract $\rho_s$ directly from the squares of spatial winding numbers, 
one has to obtain the zero-temperature values of the relevant observables.
This is computationally demanding. 
In addition, it is also crucial to include as many data points as possible
in the fits so that one can reach a accurate result of $\rho_s$. 
Hence when obtaining the ground state properties of a system is challenging, 
our approach, i.e. fitting the data using the relevant equations from 
m$\chi$PT, seems to be a more efficient way of calculating these low-energy constants since
the related simulations are performed at finite-temperature and cubical regime.

\begin{table}
\label{tab1}
\begin{center}
\begin{tabular}{cccc}
\hline
$ L_1/a $ & $c/(J_1a)$ & \\
\hline
\hline
 36 & 3.0645(11) \\
\hline
 48 & 3.0647(16) \\
\hline
 60 & 3.0651(16) \\
\hline
 72 & 3.0647(15) \\
\hline
\end{tabular}
\end{center}
\caption{The numerical values of $c$ at finite lattices for $J_2/J_1 = 1$. 
These results are obtained from the squares of spatial and temporal 
winding numbers.
}  
\end{table}

\begin{figure}
\vskip0.7cm
\begin{center}
\vbox{
\includegraphics[width=0.35\textwidth]{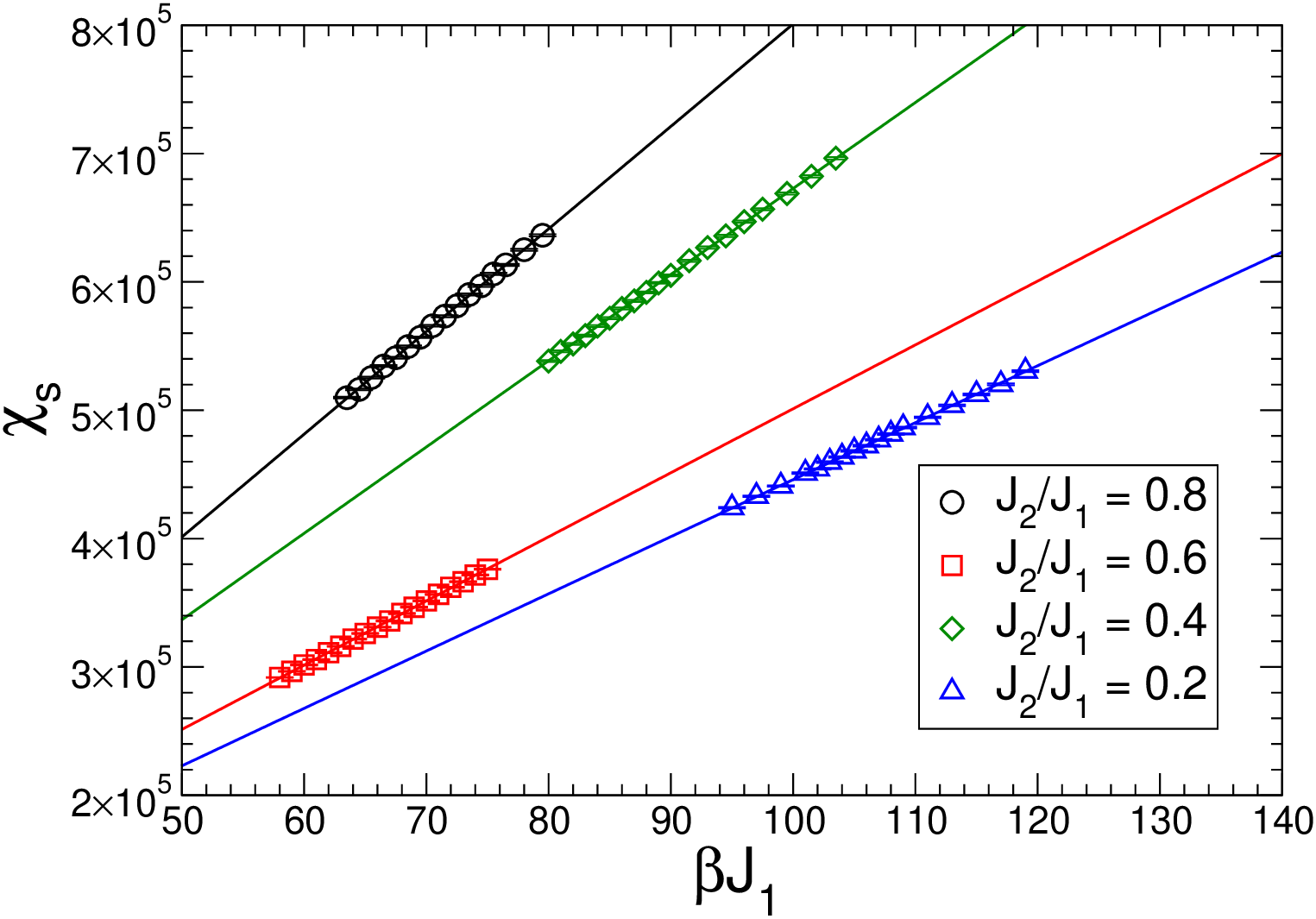}\vskip0.8cm
\includegraphics[width=0.35\textwidth]{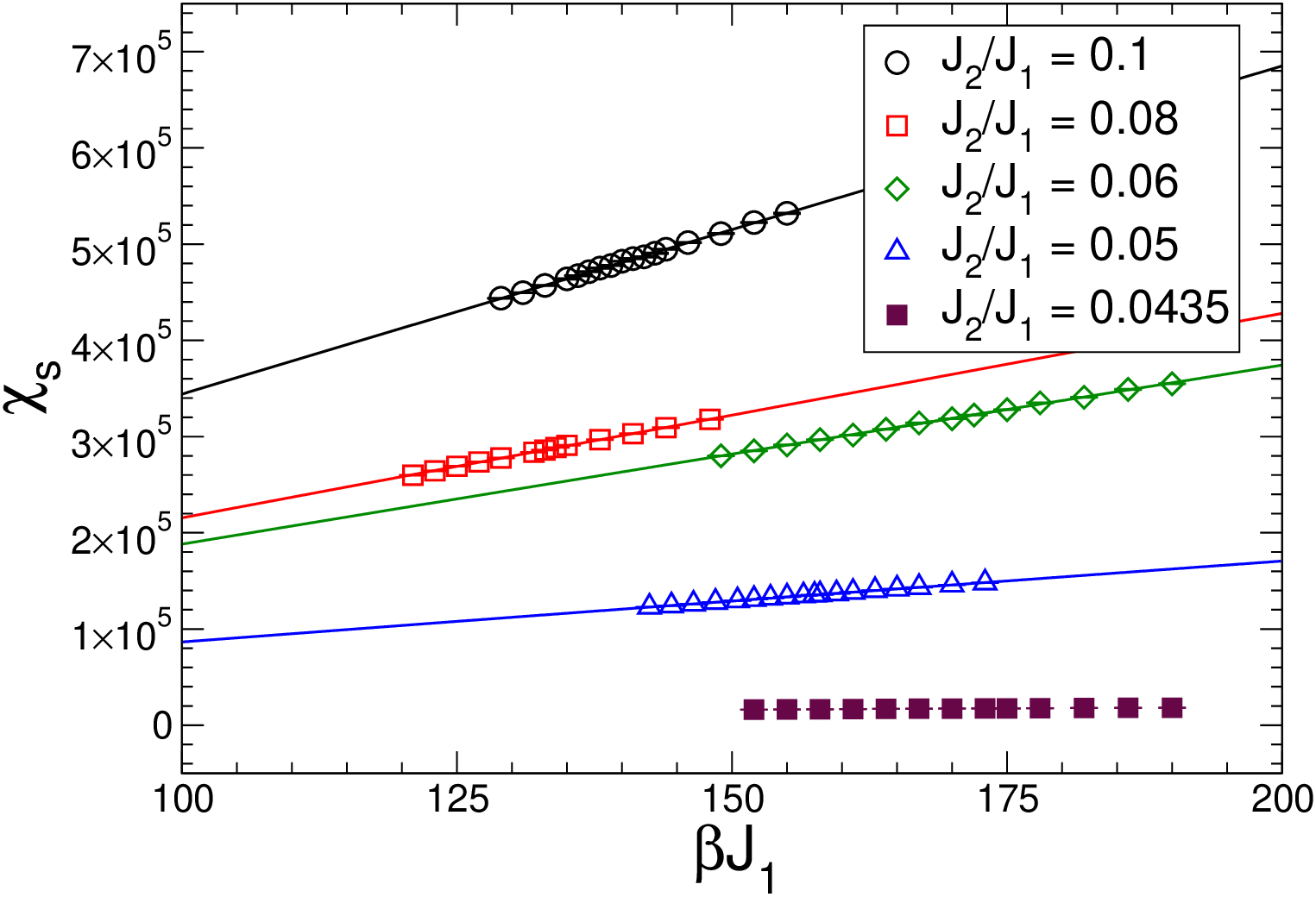}
}
\end{center}
\caption{Results of fitting the cubical regime
interpolated data points of $\chi_s$ calculated at 
$J_2/J_1$ = $0.8$, $0.6$, $0.4$, $0.2$, $0.1$, $0.08$, $0.06$, and
$0.05$ to their m$\chi$PT predictions.
The filled squares in the bottom panel are the uninterpolated 
data points determined at 
$J_2/J_1 = 0.0435$. The solid lines are obtained using the results
from the fits. No result of m$\chi$PT fit associated with $J_2/J_1 = 0.0435$ 
is shown since such a fit is of poor quality.}
\label{fig2}
\end{figure}

\begin{figure}
\vskip0.8cm
\begin{center}
\vbox{
\includegraphics[width=0.35\textwidth]{uni_sus_1_final.eps}\vskip0.8cm
\includegraphics[width=0.35\textwidth]{uni_sus_2_final.eps}
}
\end{center}
\caption{Results of fitting the cubical regime
interpolated data points of $\langle W_t^2\rangle$ calculated at 
$J_2/J_1$ = $0.8$, $0.6$, $0.4$, $0.2$, $0.1$, $0.08$, $0.06$, and
$0.05$ to their m$\chi$PT predictions.
The filled squares in the bottom panel are the uninterpolated data points 
determined at 
$J_2/J_1 = 0.0435$. The solid lines are obtained using the results from
the fits. No result of m$\chi$PT fit associated with $J_2/J_1 = 0.0435$
is shown since such a fit is of poor quality.}
\label{fig2.5}
\end{figure}

\begin{figure}
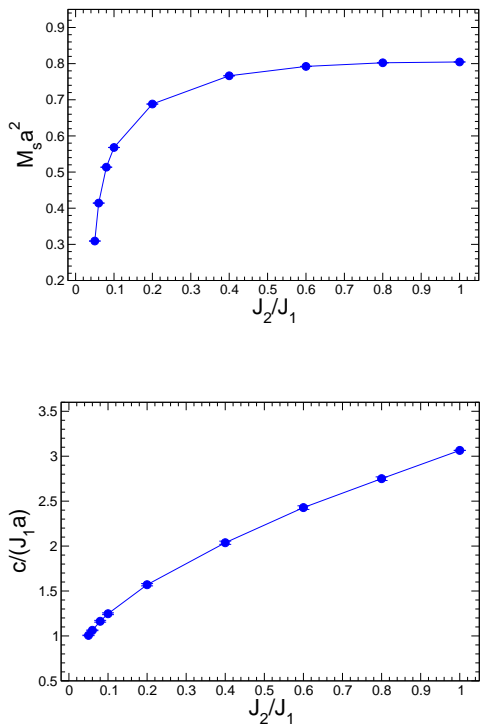

%\vskip0.0cm
\vbox{
\includegraphics[width=0.35\textwidth]{Ms_final.eps}\vskip1.0cm
\includegraphics[width=0.35\textwidth]{c_final.eps}
}
\caption{The Monte Carlo results of ${\cal M}_sa^2$ (top panel) and $c/(J_1a)$ (bottom panel)  
as functions of $J_2/J_1$. The solid lines are added to guide the eyes.}
\label{fig3}
\end{figure}

\subsection{The low-energy constants for the anisotropic models}

\begin{figure}
\vskip1.0cm
\begin{center}
\includegraphics[width=0.35\textwidth]{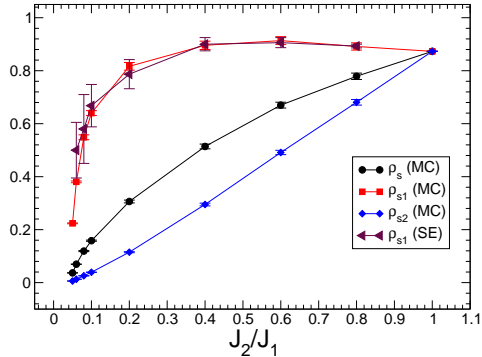}
\end{center}
\caption{The Monte Carlo as well as the series expansion results of $\rho_{s1}$,
$\rho_{s2}$, and $\rho_s$ 
as functions of $J_2/J_1$. The series expansion results
shown in the figure are estimated from Ref.~\cite{Pardini08}. The solid lines
are added to guide the eyes.}
\label{fig5}
\end{figure}

After having determined high precision values of ${\cal M}_s$,
$\rho_s$, and $c$ for $J_2/J_1 = 1.0$, we turn to the calculations of 
these low-energy constants for the anisotropic models. Similar
to the strategy used for the calculations associated with $J_2/J_1 = 1.0$, the numerical
values of $c$ for various anisotropies $J_2/J_1$ considered here are determined
using the square of winding numbers first. In particular, for each 
$J_2/J_1 \ne 1.0$ the box sizes $L_1$ and $L_2$ as well as $\beta$ 
are chosen so that the condition
$\langle W_1^2 \rangle$ $\approx$ $\langle W_2^2 \rangle$ $\approx$ 
$\langle W_t^2 \rangle$ is satisfied.
Notice interpolated data of $\chi_s$ and 
$\langle W_t^2 \rangle$, based on the spatial winding numbers squared, are used 
in order to employ Eqs.~(\ref{chiscube}) and (\ref{chiucube}) for the fits.
In addition, the effective box sizes $L$ shown in Eqs.~(\ref{chiscube}) and 
(\ref{chiucube}) are given by $L = \sqrt{L_1L_2}$.
Figs.~\ref{fig2} and \ref{fig2.5} demonstrate the results of the fits for
all the considered $J_2/J_1$. 
The obtained ${\cal M}_s$, $\rho_{s}$, and $c$
are shown in table 2, figs.~\ref{fig3} and \ref{fig5}.
Notice fig.~\ref{fig3} indicates that the antiferromagnetism is indeed weakened
as one increases the anisotropy. Furthermore, the numerical values of 
$\rho_{s1}$ obtained from our Monte Carlo data and the series expansion      
results determined in \cite{Pardini08} are in nice agreement 
from $J_2/J_1 = 1$ down to $J_2/J_1 \approx 0.08$.
Although the truncation errors of series expansion results are large
for small values of $J_2/J_1$, the outcomes of series expansion without
the truncation errors agree very well with those of Monte Carlo for $J_2/J_1 \ge 0.08$.

Interestingly, while for $J_2/J_1 \ge 0.05$ the $\chi^2/{\text{DOF}}$ of the 
fits are smaller than 1.2, the fit using the data 
of $J_2/J_1 = 0.0435$ has a very poor quality. Specifically,
we arrive at a $\chi^2/{\text{DOF}} \ge 38$ by fitting the interpolated data of 
$\chi_s$ and $\langle W_t^2\rangle$ calculated at $J_2/J_1 = 0.0435$
to Eqs.~(6) and (7). This implies that
the data of $\chi_s$ and $\langle W_t^2 \rangle$ determined at $J_2/J_1 = 0.0435$
cannot be described by Eqs.~(\ref{chiscube}) and 
(\ref{chiucube}), hence no antiferromagnetic order is present
in the system. In other words, $J_2/J_1 = 0.0435$ is
already beyond the critical point. This finding agrees with the conclusion
obtained in Ref.~\cite{Matsumoto02} that the critical point
$(J_2/J_1)_c$ is given by $(J_2/J_1)_c = 0.043648(8)$. Although
$J_2/J_1 = 0.0435$ is only slightly away from $(J_2/J_1)_c = 0.043648(8)$,
it is remarkable that
the signal for the breaking down of the long-range antiferromagnetic order
is persuasive. Interestingly,
at $J_2/J_1 = 0.0435$ the magnitude of $\chi_s$
($\langle W_t^2 \rangle$) increases (decreases) with $\beta$ (The changes are not significant). 
This is similar to their expected behavior in the N\'eel phase.
Hence, without the information of $(J_2/J_1)_c = 0.043648(8)$ and the result of
poor fitting quality,
one might naively conclude that $J_2/J_1 = 0.0435$ is still in the broken phase.
We would like to emphasize the fact that with a careful analysis using
the conventional approach, one still
reaches the same conclusion that the
long-range antiferromagnetic order is not present at $J_2/J_1 = 0.0435$. 

To employ Eqs.~(\ref{chiscube}) and 
(\ref{chiucube}) in our analysis, certain constraints such as large enough lattices must be fulfilled.
Hence one may suspect that the poor fitting quality associated with the data at $J_2/J_1 = 0.0435$
is because the required conditions for the validity of Eqs.~(\ref{chiscube}) and 
(\ref{chiucube}) are not met. To rule out this possibility, we have performed simulations
on smaller lattices for $J_2/J_1 = 0.0435$. The $\chi^2/{\text{DOF}}$ for the 
newly determined data on smaller lattices is given by $\chi^2/{\text{DOF}} \ge 3.0$.
Notice if the poor fitting quality associated with the data of $J_2/J_1 = 0.0435$ is rooted 
in the fact that the conditions for our simulations do not meet the validity requirement of Eqs.~(\ref{chiscube}) and 
(\ref{chiucube}), then the fitting results related to the data of smaller lattices should
have a worse $\chi^2/{\text{DOF}}$ than that of larger lattices. Hence, one should consider 
the poor fitting quality from the fit using the data obtained at $J_2/J_2 = 0.0435$ as
a signal for the breaking down of antiferromagnetism.     
In summary, the results demonstrated in table 2, figs. \ref{fig3}
and \ref{fig5} not only confirm the quantitative correctness of calculating the 
low-energy constants with the method employed here,
these conclusions also provide convincing evidence that the m$\chi$PT can be 
applied efficiently to detect the breaking down of the long-range 
antiferromagnetic order.

\begin{table}
\label{tab2}
\begin{center}
\begin{tabular}{ccccccc}
\hline
$J_2/J_1$ & $L_1$ & $L_2$ & ${\cal M}_sa^2$ & $\rho_s/J_1$ & $c/(J_1a)$ \\
\hline
\hline
1.0 & 100 & 100 & 0.80460(4)& 0.8731(5) & 3.0648(7)\\
\hline
0.8 & 206 & 180 & 0.8024(2) & 0.779(12) & 2.75(2)\\
\hline
0.6 & 180 & 132  & 0.7923(2) & 0.670(11) & 2.428(20)\\
\hline
0.4 & 244 & 140 & 0.7664(2) & 0.514(9)& 2.037(18)\\
\hline
0.2 & 272 & 102 & 0.6882(2) & 0.3062(54)&1.570(13)\\
\hline
0.1 & 356 & 88 & 0.5680(2) & 0.1582(23)&1.247(10)\\
\hline
0.08 & 332 & 72 & 0.5135(2) & 0.119(2)& 1.163(10)\\
\hline
0.06 &416 & 76 & 0.4142(2) & 0.0695(6)&1.0623(40)\\
\hline
0.05 & 390& 64 & 0.3092(2) & 0.03672(26)& 1.0063(32)\\
\hline
\end{tabular}
\end{center}
\caption{The numerical values of ${\cal M}_s$, $\rho_{s}$, and
$c$ determined from the fits. $\rho_{s1}$ ($\rho_{s2}$)
can be obtained by the relation $\rho_{s1} = \frac{L_1}{L_2}\rho_s$
($\rho_{s2} = \frac{L_2}{L_1}\rho_s$).
The $\chi^2/{\text{DOF}}$ for
all the considered values of $J_2/J_1$ are 
smaller than 1.2 except for $J_2/J_1 = 0.0435$
which has a $\chi^2/{\text{DOF}} \ge 38$.}  
\end{table}

\section{Conclusions and Discussions}
In this study, we have calculated the low-energy constants, namely the 
spin stiffness $\rho_s$, the staggered magnetization density 
${\cal M}_s$ per area, 
and the spinwave velocity $c$ of the spin-1 Heisenberg model with 
antiferromagnetic couplings $J_{1}$ and $J_{2}$ on the rectangular lattices using
the quantum Monte Carlo simulations. The relevant 
finite-volume and -temperature predictions of m$\chi$PT are employed in extracting the numerical values
of these low-energy constants. Such an approach is of computational efficiency as well,
since the related simulations are conducted at finite temperatures.
The precision of
${\cal M}_s$, $\rho_s$, and $c$ obtained here for $J_2/J_1 = 1.0$
is improved. 
Furthermore, the anisotropy $J_2/J_1$
dependence of ${\cal M}_s$, $\rho_{s}$ ($\rho_{s1}$), and $c$
are investigated in detail as well. In particular, the results
of ${\cal M}_s$ and $c$ determined here for the spatially anisotropic models 
are new and can serve as benchmarks for future related
studies. Our Monte Carlo and the
series expansion results of $\rho_{s1}$ are in
nice agreement for $J_2/J_1 \ge 0.08$ \cite{Pardini08}.
Consequently the quantitative correctness of our approach is justified. 
It is remarkable that the
series expansion method leads to consistent values of $\rho_{s1}$ 
with those from Monte Carlo simulations in such strong anisotropic regime 
$J_2/J_1 \ge 0.08$. We also confirm
that the m$\chi$PT can be used efficiently
to study the breaking down of long-range (antiferromagnetic) order.
Specifically, for a considered relevant parameter, 
one can conclude that a phase transition from the long-range antiferromagnetic
phase to a disordered phase 
already takes place before reaching that given parameter,   
if the fits using predictions of m$\chi$PT lead to poor fitting quality. 
Indeed, the outcomes of the fit using the data of $J_2/J_1 = 0.0435$
is consistent with the conclusion obtained in Ref.~\cite{Matsumoto02} that the 
critical point $(J_2/J_1)_c$ is given by $(J_2/J_1)_c = 0.043648(8)$. 
Considering the subtlety of quantitatively capturing the 2D characters of the spin-1/2 quantum Heisenberg 
model with the same spatial anisotropy as the one considered here, our study 
paves a way to unbiasedly investigate the novel phenomena of 2D to 1D dimension 
crossover of the related spin-1/2 model. 

%\vskip2cm
~
~
~
~
~
~
~
~
~
~

\section{Acknowledgments}
%\vskip-0.25cm
Simulations were done based on the loop algorithms in ALPS \cite{Troyer08,Bau11}.
F.-J.~J. is supported partially by MOST of Taiwan.

\end{document}